\documentclass[aps,prl,twocolumn,showpacs]{revtex4}
\usepackage{amssymb}
\usepackage{amsmath}
\usepackage{graphicx}

\begin{document}
\title{Observation of quantum jumps of a single quantum dot spin using sub-microsecond single-shot optical readout}
\author{Aymeric Delteil}
\author{Wei-bo Gao}
\author{Parisa Fallahi}
\author{Javier Miguel-Sanchez}
\author{Atac Imamo\u{g}lu}

\affiliation{Institute of Quantum Electronics, ETH Zurich, CH-8093
Zurich, Switzerland.}

\date{\today }

\begin{abstract}

Single-shot read-out of individual qubits is typically  the slowest
process among the elementary single- and two-qubit operations
required for quantum information processing. Here, we use resonance
fluorescence from a single-electron charged quantum dot to read-out
the spin-qubit state in $800$ nanoseconds with a fidelity exceeding
$80 \%$. Observation of the spin evolution on longer timescales
reveals quantum jumps of the spin state: we use the experimentally
determined waiting-time distribution to characterize the quantum
jumps.

\end{abstract}

 \maketitle

A fundamental difficulty in quantum information processing is the
need for isolation of individual quantum systems from their noisy
environment on the one hand, and the requirement for information
extraction by selective coupling of qubits to classical (noisy)
detectors on the other hand \cite{Divincenzo}. The requisite one-
and two-qubit operations, as well as initialization of each qubit
can be carried out by using classical out-of-equilibrium external
fields, such as lasers or microwaves; the lack of a need for
heralding the successful completion of these operations ensures that
they can be accomplished in short timescales. In contrast, quantum
measurements are typically slow since information extraction by a
classical observer is in many cases hindered by the need to protect
the qubit from the external fluctuations. While ingenious schemes
for fast qubit measurements have been developed, the timescales
required for a high fidelity qubit measurement remains at least an
order of magnitude longer than those required for coherent
operations in practically all quantum information processing schemes
\cite{Elzerman04, Morello10,Robledo11}. In the case of spin qubits
in optically active quantum dots (QD), the predicament is even more
striking since while optical excitation allows for fast turn on/off
of light-matter interaction enabling spin read-out, it at the same
time allows for an additional fast channel for spin relaxation. In
fact, with the exception of a slow coupled QD scheme requiring a
designated read-out QD~\cite{Vamivakas10}, it has not been possible
to carry out single-shot spin measurements on isolated optically
active spin qubits~\cite{Chaoyang10}.

In this Letter, we overcome the predicament underlying single-shot
spin read-out by enhancing the collection efficiency of resonance
fluorescence (RF) from spin-dependent recycling transitions that are
ubiquitous to single-electron charged QDs. The photon collection
efficiency of $0.45 \%$ that we achieve allows us to obtain a
single-shot spin read-out fidelity exceeding $80 \%$ in a
measurement time of $800$~ns. This result corresponds to an
enhancement of the spin read-out time by almost three orders of
magnitude as compared to the prior measurements on coupled QDs
\cite{Vamivakas10}. Continuous monitoring of the spin state enabled
by single-shot read-out reveals quantum jumps of the observed spin
stemming either from the finite $T_1$ spin lifetime or spin pumping
induced by the resonant read-out laser. A theoretical analysis of
quantum jumps using the waiting time distribution ($W(\tau)$) was
presented earlier~\cite{Cohen86,Zoller87}. Here we use  the experimentally determined $W(\tau)$ and the second order
correlation function ($g^{(2)}(\tau)$) of the RF events to
characterize the (incoherent) spin dynamics.

Our experiment is carried out on a single InGaAs self-assembled
quantum dot.  The QD is placed in a low quality factor ($Q < 10$)
microcavity, consisting of a $28$ layer distributed Bragg reflector
(DBR) mirror underneath the dot layer and a thin metal gate with a
power reflectance $R< 0.5$ deposited on the top surface; this
structure reduces the solid-angle into which QD photons are emitted.
A solid immersion lens is mounted on the sample in order to further
increase the extraction efficiency  into a $NA =0.65$ objective to
$7.0 \%$ and the overall detection efficiency to $\sim 0.45 \%$. The
semi-transparent metallic top gate and a back $n$-doped layer form a
Schottky diode structure, which is used to control the charge state
of our quantum dot. The sample is in a liquid helium bath cryostat
with an external magnetic field applied perpendicular to the growth
direction (Faraday geometry). In this configuration, $|\uparrow\rangle\leftrightarrow|T_b\rangle$ and $|\downarrow\rangle\leftrightarrow|T_r\rangle$ are two transitions with strong oscillation strength, while the diagonal transitions $|\uparrow\rangle\leftrightarrow|T_r\rangle$ and $|\downarrow\rangle\leftrightarrow|T_b\rangle$ are only weakly allowed by the heavy-light hole mixing and have a $\sim 450$ weaker oscillation strength (Fig.~1c). 
A confocal
microscope is used to focus the excitation laser on the quantum dot
as well as to collect photons from the quantum dot. In the
excitation and collection arms, cross polarization technique
\cite{Vamivakas10,Yilmaz10} is used to suppress the reflected laser
background by a factor $\sim 10^{-6}$. Scattered photons from the
quantum dot are channeled to a superconducting  single-photon
detector (SSPD) and the detection events are analyzed using a
time-correlated single photon counting module.

To fully characterize the QD, we perform a two-color RF measurement
in the single-electron charged regime using the pulse sequence shown
in Fig.~1a at $B = 2T$. The two laser pulses with duration time
$3.8~\mu s$, separated by an interval of $0.5~\mu s$, are generated from continuous-wave (cw) lasers using
amplitude electro-optic modulators (EOM) driven by two synchronized
pulse pattern generators. We fix the wavelength of laser 2 to
$961.795~nm$ and measure RF as a function of the gate voltage and the
wavelength of laser 1 (Fig.~1b). In the center of the plateau, the
RF signal disappears due to spin pumping \cite{Atature06,Xu07}, with
the exception of two particular wavelengths (marked with red lines
$c$ and $d$) where the signal is recovered due to spin re-pumping:
in these two cases, at gate voltage $V = 0.235$~V, laser 2 is
resonant with the red vertical transition, and laser 1 is resonant
with either the blue vertical transition or the diagonal transition
originating from $| \uparrow \rangle$, as shown in the corresponding
energy level diagrams in Fig.~1c and Fig.~1d. With the two
successive pulses applied on the quantum dot, the spin is pumped
back and forth between $|\uparrow\rangle$ and $|\downarrow\rangle$
states, ensuring that the RF signal is recovered.  The different
lineshapes observed in Fig.~1b in these two cases are most likely
due to dynamic nuclear spin polarization effects
\cite{LattaNuclear}. Using resonant cw excitation  at zero magnetic
field we detect $2.6$ million counts per second from the QD with
excitation laser power above QD saturation. The trion lifetime is
$0.65ns$, i.e. the QD emits $\sim7.7\times10^8$ photons per second
when driven well above saturation. 
Taking into account the effect of the $\sim 90$~ns dead time of our time-correlated single photon counting module, our overall collection
efficiency is $0.45\%$. After correcting for the SSPD efficiency
($\simeq 40\%$), beam splitter ($80\%$) and polarizer ($50\%$)
losses as well as the fiber coupling efficiency ($40\%$), we
conclude that $7.0\%$ of the photons emitted by the QD are collected
by the objective.

\begin{figure}[h]
  \centering
  \includegraphics[width=3.5in]{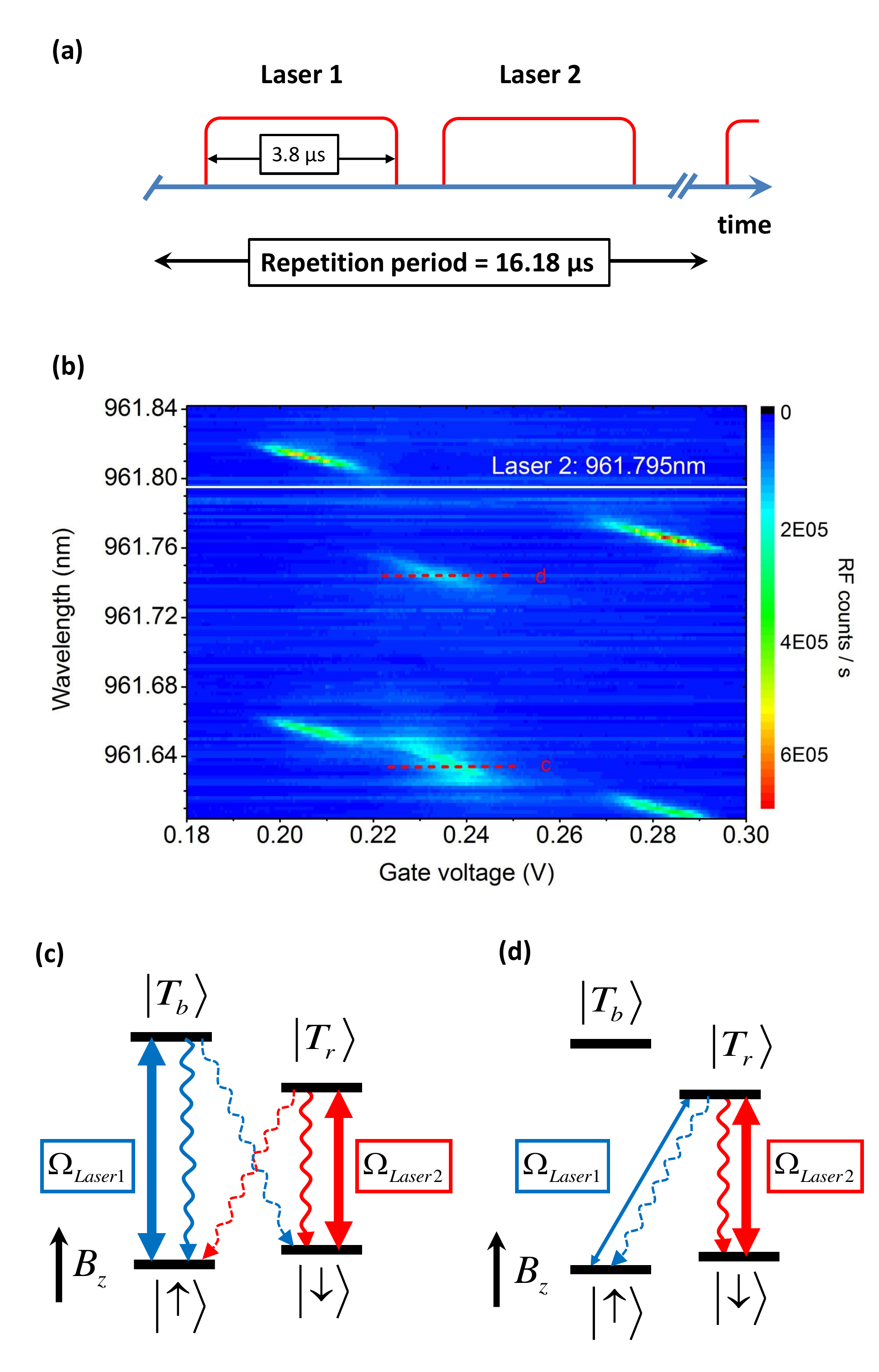}\\
  \caption{(a) Pulse sequence used in the plateau scan as well as in the single-shot experiment. The two laser pulses have the same time duration. (b) The resonance fluorescence (RF) counts as a function of gate voltage and laser wavelength of laser 1. Here the RF counts include the total counts detected during both pulses. The magnetic field we use is $2T$ and the wavelength of laser 2 is fixed at $961.795nm$. In the middle of the plateau, the signal disappears due to spin pumping except when the laser 1 wavelength is at $961.635nm$ or $961.745nm$ (marked with red lines) at a gate voltage $\sim 0.236V$. At these two wavelengths, the corresponding energy level diagram is shown in (c) and (d) respectively.  }\label{1}
\end{figure}

Figure~2 summarizes the experiments  demonstrating single-shot spin
readout on sub-microsecond time-scales.  With the pulse sequence
depicted in Fig.~1, we record time resolved RF from the QD.
Figure~2a shows the average RF counts which decay exponentially
during each pulse due to spin pumping \cite{Atature06}. By comparing
the counts at the beginning and the end of laser 1 (2) pulse that is
resonant with the vertical blue (red) transition, we estimate a
lower bound of the spin pumping fidelity $96.1\%\pm 1.0\% $
($96.0\%\pm 0.7\% $) for $|\uparrow\rangle$ ($|\downarrow\rangle$)
state. The imperfect spin-pumping is mostly due to  the off-resonant
excitation of the other vertical transition and the finite pulse
duration.

We remark that the detected RF signal in  the second pulse for
measuring $|\downarrow\rangle$ is higher than that in the first
pulse as a consequence of the polarization settings in our system.
In Faraday geometry, the two vertical cycling transitions
$|\downarrow\rangle\leftrightarrow|T_r\rangle$ and
$|\uparrow\rangle\leftrightarrow|T_b\rangle$ have equal oscillator
strengths and are $\sigma^-$ and $\sigma^+$ circularly polarized,
respectively. We set the polarization of both lasers to be $\alpha
\sigma^+ + \beta \sigma^-$. To suppress the reflected laser
background, the polarizer in the collection arm is set so that it
transmits $\alpha^* \sigma^- - \beta^* \sigma^+$ polarized light. In
the limit $|\alpha|>|\beta|$, the detected $\sigma^-$ RF counts will
be larger by a factor $|\alpha|^2/|\beta|^2$, provided that we
increase the intensity of laser 2 to ensure that the
$|\downarrow\rangle\leftrightarrow|T_r\rangle$ transition is driven
to saturation. From the total photon numbers in each pulse, we
obtain $\alpha^2/\beta^2= 2.60\pm0.01$ for our experiment.


Before describing our principal experimental observations, we argue
that a natural definition of a single-shot measurement is provided
by a comparison between the average waiting time $t_{wait}$ between
two successive photon detection events and the spin-flip time $t_{spin}$.
We refer to the read-out procedure as {\sl single-shot} if $t_{wait}
\le t_{spin}$: in this limit we detect (on average) $\ge 1$ photons before the spin flips from the bright state to the dark state
~\footnote{Note that the
presence of a dead time in the detection apparatus will decrease the
average photon number but not the measurement fidelity as it affects
only the situations involving at least one detection event.}. As we
discuss shortly, the time constants $t_{wait}$ and $t_{spin}$ emerge
naturally in waiting time distribution $W(\tau)$ and second order correlation function $g^{(2)}(\tau)$.


To perform a single-shot spin measurement, we prepare the spin state
with laser 1 either in $|\uparrow\rangle$ or $|\downarrow\rangle$;
afterwards, the spin state $|\downarrow\rangle$ is read out with
laser 2 that is kept on resonance with
$|\downarrow\rangle\leftrightarrow|T_r\rangle$ and its power is
chosen to be $22nW$, about 4 times the saturation power (Fig.~2b).
Figure~2c shows the probability distribution of the photon counts in
the first $800ns$ of the read-out pulse, corresponding to the grey
time window in Fig.~2a. When the spin state is prepared by laser 1
in $|\uparrow\rangle$, detecting zero photons is the most likely
outcome. Conversely, for the spin prepared in $|\downarrow\rangle$,
it is more likely that one or more photons are detected. We find
that in this latter case, the average number of detected photons is
$1.27\pm 0.01$, demonstrating single-shot measurement of the
electron spin state. The deviation of the detected photon number
distribution from a geometric distribution is mainly due to the dead
time of our time-correlated single photon counting module
($\sim90~ns$) and the size of the detection window which is
comparable to the spin lifetime.

When the initial state is $|\downarrow\rangle$, the  average
detected photon number $\langle n \rangle$ increases with a larger
detection window, mainly due to incomplete spin pumping in the first
$800ns$ depicted in Fig.~2c. For a detection window of $3.8\mu s$,
$\langle n \rangle$ is increased to $\sim 2$ (Fig.~2d). For
$|\uparrow\rangle$, the average counts will also increase with
increasing measurement/recording time to ~0.1; these counts stem
primarily from the residual reflected laser photons. In the spin
state measurement, if we detect no photons, we assign the spin state
as $|\uparrow\rangle$. If on the other hand, we detect one or more
photons, we assign the spin state prior to the measurement pulse as
$|\downarrow\rangle$. The average spin read-out fidelity
$F_{avg}=1/2(p_{|\uparrow\rangle}+p_{|\downarrow\rangle})$ we
measure is $0.823\pm0.002$ for a detection window of $800ns$ and
$0.826\pm0.002$ for a detection window of $3.8\mu s$. Here
$p_{|\uparrow\rangle}$ is the probability of detecting no photons
when the initial spin state is $|\uparrow\rangle$;
$p_{|\downarrow\rangle}$ denotes the probability that at least one
photon is detected when the initial state is $|\downarrow\rangle$).

\begin{figure}[h]
  \centering
  \includegraphics[width=3.5in]{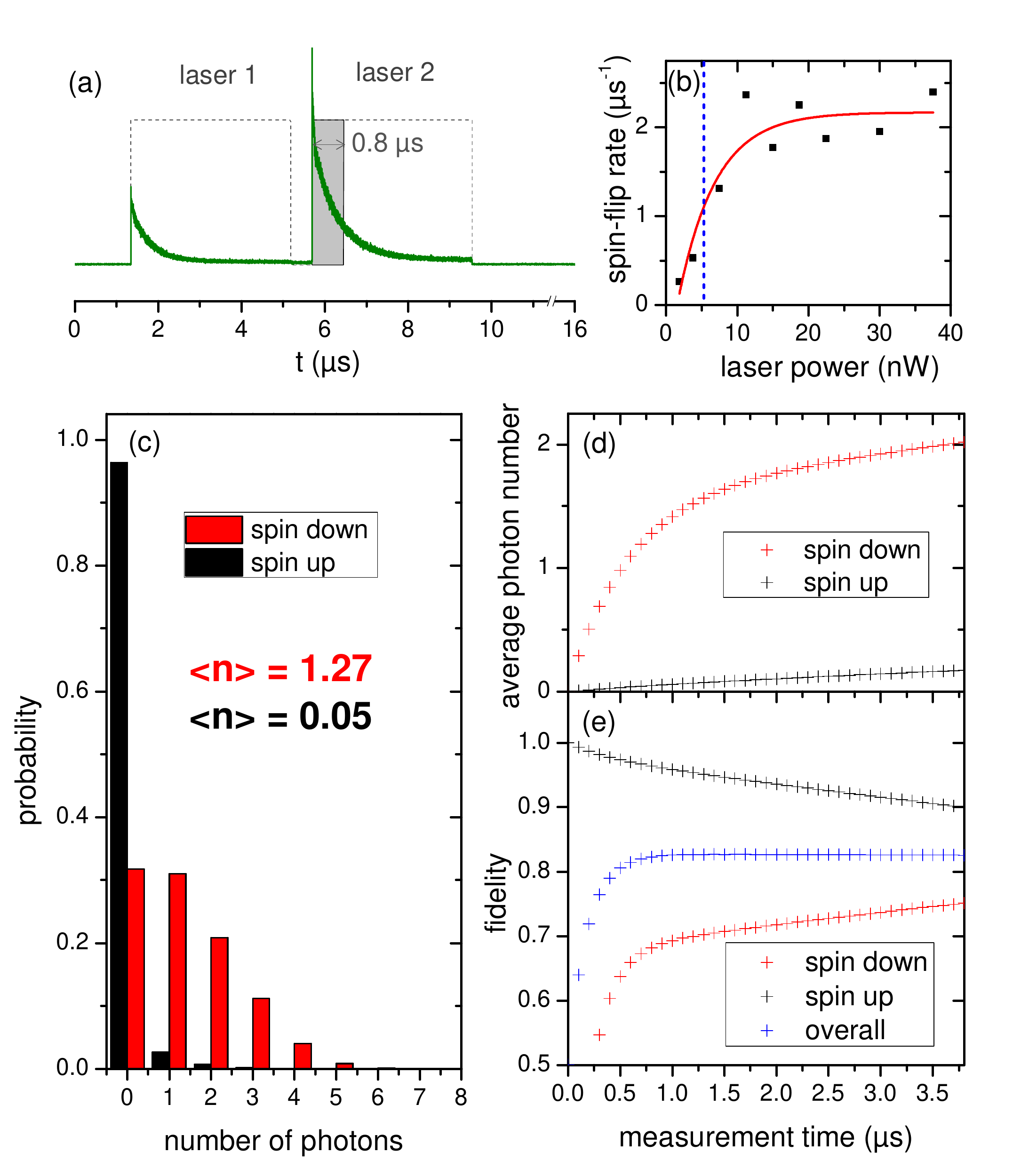}\\
  \caption{(a) The time resolved average RF counts measured with two lasers as shown in the energy level diagram in Fig.~1(c). The time range used to calculate the photon number probability in (c) is shown in grey area. (b) Spin flip rate as a function of the excitation laser power in the readout pulse. The saturation power is indicated by the dashed blue line. The error bars are smaller than the symbols, and the deviation from the exponential shape (red line) comes from laser-power-dependent dynamical nuclear spin polarization effects. (c) The normalized probability of photon number detected in a time range of $800ns$ after the spin is prepared to the $|\uparrow\rangle$ (black column) or $|\downarrow\rangle$ (red column). The statistics is  obtained from data in $10$ seconds (618000 repetitions). (d) The average number of detected photons for spin up and spin down state as a function of the readout duration. (e) The probability of detecting spin up and spin down, and the overall fidelity as a function of the read-out duration. The error bar is smaller than the symbols.}\label{1}
\end{figure}

Single-shot read-out capability enables the observation of quantum
jumps in spin dynamics. In our experiments, the changes in the spin
state are predominantly due to spin-flip Raman scattering processes.
To observe the associated spin jumps, we choose a two laser
excitation configuration depicted in Fig.~1d. A strong cw laser
resonant with the red vertical transition is used to detect the
$|\downarrow\rangle$ state while inducing spin pumping into
$|\uparrow\rangle$. A second cw laser resonant with the diagonal
transition is used for inducing spin flips from $|\uparrow\rangle$
back to $|\downarrow\rangle$. The intensity of the lasers are chosen
to ensure a spin pumping (repumping) time $ \sim 1~\mu s$ ($\sim
10~\mu$s). The photon detection events are shown in Fig.~3a for a
detection time window of $100~\mu$s, for two different values $P_a$
and $P_b$ of the repumping laser power: the registred detection
events  bunch together into separated clusters, showing alternating
bright and dark periods that indicate jumps in the electron spin
state.

In order to extract the characteristic constants of the spin
dynamics from the continuous measurement time traces, we calculate
the functions $g^{(2)}(\tau)$  (second order correlation function) and $W(\tau)$ (waiting time distribution) from the experimental
data. The unnormalized $g^{(2)}(\tau)$ curve, obtained from $1$~s
long RF traces, is shown in Fig.~3c.  The bunching behavior reveals
information about the spin-flip dynamics. An exponential decay fit
to the $g^{(2)}(\tau)$ curve gives a spin lifetime of $t_{spin} =
972\pm4~ns$.  Using the same data, we also determine $W(\tau)$ which
gives, conditional on detecting one photon, the probability of
detecting a second photon after a waiting time $\tau$ without any detection event in between (red curve in
Fig. 3c). The construction of $g^2(\tau)$ and $W(\tau)$ is discussed in more detail in the supplementary material. In the inset of the Fig. 3c, the normalized $W(\tau)$ is
shown in log-scale for two different values of the diagonal
repumping laser power. Two exponential decays can be
observed~\cite{Cohen86,Zoller87}: the first one with a time-constant
$t_{wait}$ stems from the detection of the second photon while the
spin state remains the same. The much longer second decay time
($t_{repump}$) originates from the cases where two consecutive spin
flips take place in between two detection events; thus, $t_{repump}$
strongly depends on the repumping laser power whereas $t_{wait}$ is
independent of it. This allows to write $W(\tau)$ as the sum of
these two exponential contributions, namely
$W(\tau)=W_{short}(\tau)+W_{long}(\tau)$, where $W_{short}(\tau)$
($W_{long}(\tau)$) is the component with the short (long) decay time
constant $t_{wait}$ ($t_{repump}$). The low value of both
$g^{(2)}(\tau=0)$ and $W(\tau=0)$ is caused by the dead time of our
time-correlated single photon counting module and does
not correspond to the actual physical value of these functions for
$\tau =0$. The fact that $W(\tau)$ does not present sizable
deviation from the biexponential shape indicates that the dynamics
of our system in the considered time scales can be treated
incoherently using rate equations~\cite{Zoller87}.

The three time constants, namely $t_{spin}$, $t_{wait}$ and
$t_{repump}$, are the relevant quantities in the following analysis
of quantum jumps. The observation of quantum jumps requires the
fulfillment of two conditions: first, $t_{spin}$ should be longer
than $t_{wait}$ in order to detect a significant number of photons
when the spin is in the $|\downarrow\rangle$ state. This condition
is equivalent to the single-shot read-out condition, which is
satisfied in our experiments. Second, $t_{repump}$ has to be much
longer than $t_{wait}$ in order to make a clear distinction between
waiting events occurring while the spin remains in the
$|\downarrow\rangle$ state, and longer waiting events associated
with two consecutive spin-flip process. These conditions lead to
binary RF signal, presenting alternating bright and dark periods
with abrupt changes. From the fits depicted in Fig.~3c, we find in
our case $t_{repump}>t_{spin}>t_{wait}$, allowing the observation of
quantum jumps. To identify a waiting period $(t_1,t_1+\tau)$ between
two consecutive photon detection events as bright or dark, we use
$W(\tau)$: if $W_{short}(\tau)>W_{long}(\tau)$
($W_{short}(\tau)<W_{long}(\tau)$), then we identify the period
$(t_1,t_1+\tau)$ as a bright (dark) period and shade it in Fig.~3a
in grey (white).

A direct distinction between these  waiting events of very different
origins can be made as well by binning the data with a judicious
choice of the bin size $T_{bin}$, such that $t_{repump} \gg T_{bin}
> t_{wait}$. In this case, the short waiting events are integrated
in the bins: a change in $n$ for two consecutive bins from $n\geq 1$
to $n =0$ identifies a quantum jump. Fig.~3b presents the same data
as in the lower panel of Fig.~3a, with the counts integrated into
$1~\mu s$ bins. The inferred bright periods are in excellent
agreement with the identification based on comparing
$W_{short}(\tau)$ and $W_{long}(\tau)$, depicted in the lower panel
of Fig.~3a. The dark periods dominate over the bright periods since
we have set $t_{repump} > t_{spin}$.

\begin{figure}[h]
  \centering
  \includegraphics[width=3in]{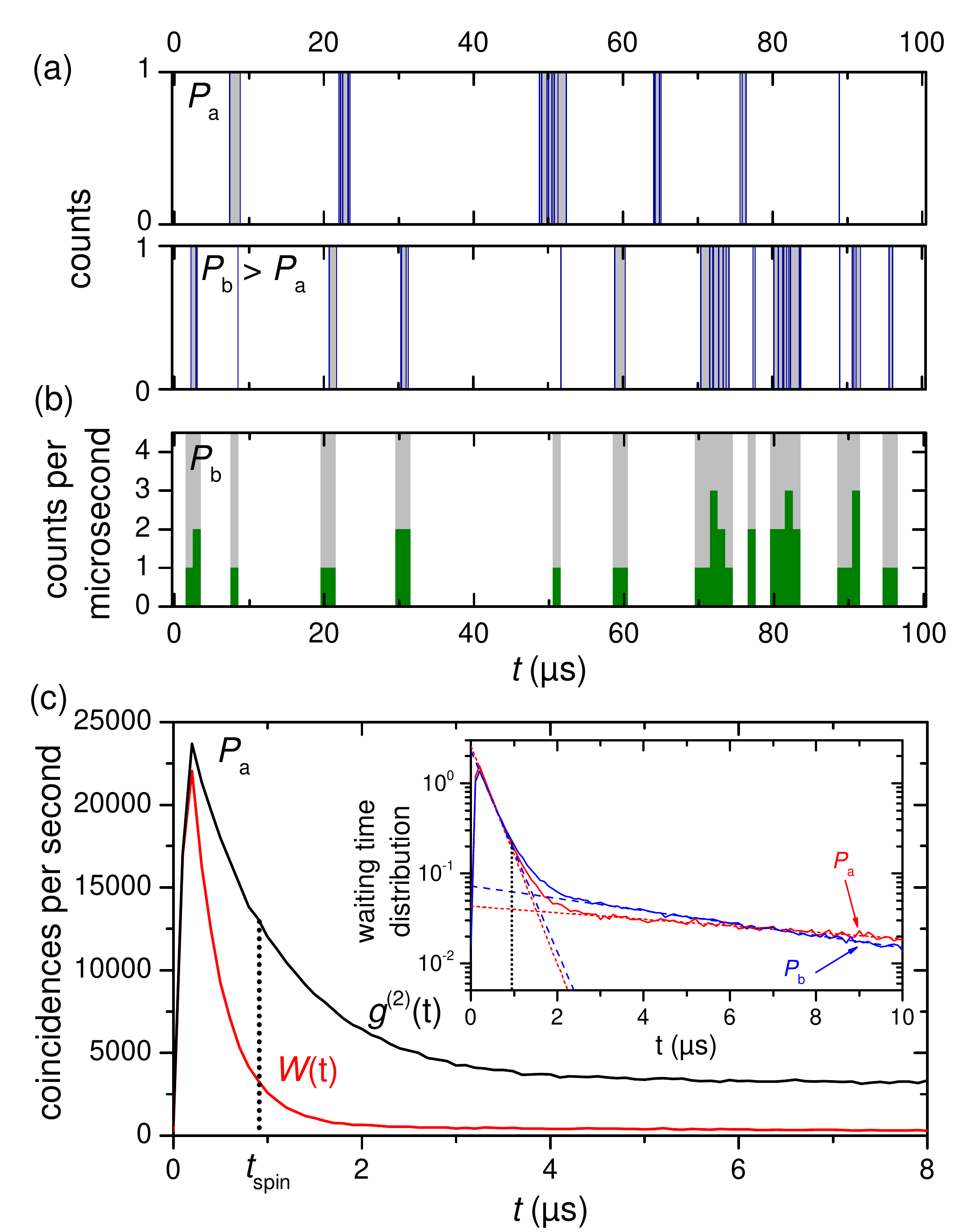}\\
  \caption{(a) Quantum jumps in continuous readout. The power of laser used for exciting the diagonal transition is weaker for the upper panel ($P_a$) and stronger for  the lower panel ($P_b > P_a$). (b) Same data as (a), lower panel, but the counts are stored in 1$\mu s$ bins. (c)  Second order correlation function $g^{(2)}(t)$ (black trace) and waiting time distribution (red trace) for the data in the upper panel of (a). The dashed line shows the spin lifetime.  Inset: the normalized waiting time function for the upper panel  (red line) and the lower panel (blue line) in (a). Two exponential decays are observed.  The fitted decay times for the red line are $t_{wait}=349 \pm 5 ns$ and $t_{repump}= 12.1\pm 0.3 \mu s$ and for the blue line $t_{wait}=371 \pm 5 ns$ and $t_{repump}= 6.32\pm 0.07 \mu s$ }\label{1}
\end{figure}

In summary, we report the observation of sub-microsecond all-optical
single-shot measurement of an isolated electron spin confined in a
single quantum dot. We expect this result to stimulate research
aimed at probabilistic entanglement of distant spins which have so
far been hindered by inefficient multi-shot spin measurements.
Embedding a quantum dot in a photonic nanostructure could be used to
enhance the collection efficiency by a factor of 10 \cite{Gazzano13}.
Together with a Purcell enhancement factor  $F_p \sim 4$
\cite{Gazzano13}, the measurement could be achieved within
$20$~nanoseconds with a fidelity of 95~\%. We emphasize in addition that the characterization of spin jumps using the waiting time distribution shows the power of
quantum optical measurements in identifying the elementary
properties of optically active solid-state qubits.

This work is supported by NCCR Quantum Science and Technology (NCCR QSIT), research instrument of the Swiss National Science Foundation (SNSF), by Swiss NSF under Grant No. 200021-140818, an ERC Advanced Investigator Grant (A.I.). The research leading to these results has received funding from the European Union Seventh Framework Programme (FP7/2007-2013) under grant agreement n$^{o}$ 289795. A.D. and W.-B.G. contributed equally to this work.

\end{document}